\documentclass[ALICE,manyauthors]{cernphprep}
\usepackage[comma,square,numbers,sort&compress]{natbib}
\usepackage{hyperref}
\usepackage{lineno}
\usepackage{xspace}
\usepackage{xcolor}
\usepackage{paralist}
\usepackage[T1]{fontenc}
\usepackage{orcidlink}

\begin{document}
%

\newcommand{\pp}           {pp\xspace}
\newcommand{\ppbar}        {\mbox{$\mathrm {p\overline{p}}$}\xspace}
\newcommand{\XeXe}         {\mbox{Xe--Xe}\xspace}
\newcommand{\PbPb}         {\mbox{Pb--Pb}\xspace}
\newcommand{\pA}           {\mbox{pA}\xspace}
\newcommand{\pPb}          {\mbox{p--Pb}\xspace}
\newcommand{\AuAu}         {\mbox{Au--Au}\xspace}
\newcommand{\dAu}          {\mbox{d--Au}\xspace}

\newcommand{\s}            {\ensuremath{\sqrt{s}}\xspace}
\newcommand{\snn}          {\ensuremath{\sqrt{s_{\mathrm{NN}}}}\xspace}
\newcommand{\pt}           {\ensuremath{p_{\rm T}}\xspace}
\newcommand{\meanpt}       {$\langle p_{\mathrm{T}}\rangle$\xspace}
\newcommand{\ycms}         {\ensuremath{y_{\rm CMS}}\xspace}
\newcommand{\ylab}         {\ensuremath{y_{\rm lab}}\xspace}
\newcommand{\etarange}[1]  {\mbox{$\left | \eta \right |~<~#1$}}
\newcommand{\yrange}[1]    {\mbox{$\left | y \right |~<~#1$}}
\newcommand{\dndy}         {\ensuremath{\mathrm{d}N_\mathrm{ch}/\mathrm{d}y}\xspace}
\newcommand{\dndeta}       {\ensuremath{\mathrm{d}N_\mathrm{ch}/\mathrm{d}\eta}\xspace}
\newcommand{\avdndeta}     {\ensuremath{\langle\dndeta\rangle}\xspace}
\newcommand{\dNdy}         {\ensuremath{\mathrm{d}N_\mathrm{ch}/\mathrm{d}y}\xspace}
\newcommand{\Npart}        {\ensuremath{N_\mathrm{part}}\xspace}
\newcommand{\Ncoll}        {\ensuremath{N_\mathrm{coll}}\xspace}
\newcommand{\dEdx}         {\ensuremath{\textrm{d}E/\textrm{d}x}\xspace}
\newcommand{\RpPb}         {\ensuremath{R_{\rm pPb}}\xspace}

\newcommand{\nineH}        {$\sqrt{s}~=~0.9$~Te\kern-.1emV\xspace}
\newcommand{\seven}        {$\sqrt{s}~=~7$~Te\kern-.1emV\xspace}
\newcommand{\twoH}         {$\sqrt{s}~=~0.2$~Te\kern-.1emV\xspace}
\newcommand{\twosevensix}  {$\sqrt{s}~=~2.76$~Te\kern-.1emV\xspace}
\newcommand{\five}         {$\sqrt{s}~=~5.02$~Te\kern-.1emV\xspace}
\newcommand{\twosevensixnn}{$\sqrt{s_{\mathrm{NN}}}~=~2.76$~Te\kern-.1emV\xspace}
\newcommand{\fivenn}       {$\sqrt{s_{\mathrm{NN}}}~=~5.02$~Te\kern-.1emV\xspace}
\newcommand{\LT}           {L{\'e}vy-Tsallis\xspace}
\newcommand{\GeVc}         {Ge\kern-.1emV/$c$\xspace}
\newcommand{\MeVc}         {Me\kern-.1emV/$c$\xspace}
\newcommand{\TeV}          {Te\kern-.1emV\xspace}
\newcommand{\GeV}          {Ge\kern-.1emV\xspace}
\newcommand{\MeV}          {Me\kern-.1emV\xspace}
\newcommand{\GeVmass}      {Ge\kern-.2emV/$c^2$\xspace}
\newcommand{\MeVmass}      {Me\kern-.2emV/$c^2$\xspace}
\newcommand{\lumi}         {\ensuremath{\mathcal{L}}\xspace}

\newcommand{\ITS}          {\rm{ITS}\xspace}
\newcommand{\TOF}          {\rm{TOF}\xspace}
\newcommand{\ZDC}          {\rm{ZDC}\xspace}
\newcommand{\ZDCs}         {\rm{ZDCs}\xspace}
\newcommand{\ZNA}          {\rm{ZNA}\xspace}
\newcommand{\ZNC}          {\rm{ZNC}\xspace}
\newcommand{\SPD}          {\rm{SPD}\xspace}
\newcommand{\SDD}          {\rm{SDD}\xspace}
\newcommand{\SSD}          {\rm{SSD}\xspace}
\newcommand{\TPC}          {\rm{TPC}\xspace}
\newcommand{\TRD}          {\rm{TRD}\xspace}
\newcommand{\VZERO}        {\rm{V0}\xspace}
\newcommand{\VZEROA}       {\rm{V0A}\xspace}
\newcommand{\VZEROC}       {\rm{V0C}\xspace}
\newcommand{\Vdecay} 	   {\ensuremath{V^{0}}\xspace}

\newcommand{\ee}           {\ensuremath{e^{+}e^{-}}} 
\newcommand{\pip}          {\ensuremath{\pi^{+}}\xspace}
\newcommand{\pim}          {\ensuremath{\pi^{-}}\xspace}
\newcommand{\kap}          {\ensuremath{\rm{K}^{+}}\xspace}
\newcommand{\kam}          {\ensuremath{\rm{K}^{-}}\xspace}
\newcommand{\pbar}         {\ensuremath{\rm\overline{p}}\xspace}
\newcommand{\kzero}        {\ensuremath{{\rm K}^{0}_{\rm{S}}}\xspace}
\newcommand{\lmb}          {\ensuremath{\Lambda}\xspace}
\newcommand{\almb}         {\ensuremath{\overline{\Lambda}}\xspace}
\newcommand{\Om}           {\ensuremath{\Omega^-}\xspace}
\newcommand{\Mo}           {\ensuremath{\overline{\Omega}^+}\xspace}
\newcommand{\X}            {\ensuremath{\Xi^-}\xspace}
\newcommand{\Ix}           {\ensuremath{\overline{\Xi}^+}\xspace}
\newcommand{\Xis}          {\ensuremath{\Xi^{\pm}}\xspace}
\newcommand{\Oms}          {\ensuremath{\Omega^{\pm}}\xspace}
\newcommand{\degree}       {\ensuremath{^{\rm o}}\xspace}

\newcommand{\angantyr}     {PYTHIA 8.3/Angantyr\xspace}

\begin{titlepage}
\PHyear{2025}       
\PHnumber{073}      
\PHdate{26 March}  

\title{Centrality dependence of charged-particle pseudorapidity density at midrapidity in Pb--Pb collisions at $\mathbf{\sqrt{\textit{s}_{\textup{NN}}} = 5.36}$ TeV}
\ShortTitle{Charged-particle multiplicity in Pb--Pb collisions} 

\Collaboration{ALICE Collaboration\thanks{See Appendix~\ref{app:collab} for the list of collaboration members}}
\ShortAuthor{ALICE Collaboration} 

\begin{abstract}
The ALICE Collaboration reports its first LHC Run 3 measurements of charged-particle pseudorapidity density at midrapidity in Pb--Pb collisions at a centre-of-mass energy per nucleon pair of $\sqrt{s_{\mathrm{NN}}}=5.36$ TeV. 
Particle multiplicity in high-energy collisions characterises the system geometry, constrains particle-production mechanisms, and is used to estimate initial energy density.
Multiplicity also acts as a reference for subsequent measurements as a function of centrality.
In this letter, for the first time, charged particles are reconstructed using the upgraded ALICE Inner Tracking System and Time Projection Chamber, while the collision centrality is determined by measuring charged-particle multiplicities with the Fast Interaction Trigger system. 
Pseudorapidity density, \dndeta, is presented, averaged over events, for various centrality classes. 
Results are shown as a function of pseudorapidity and the average number of participating nucleons ($\langle N_{\mathrm{part}}\rangle$) in the collision.
The average charged-particle pseudorapidity density (\avdndeta) at midrapidity ($|\eta|<0.5$) is 2010 $\pm$ 53 for the 5\% most central collisions.
The value of \avdndeta normalised to $\langle N_{\mathrm{part}}\rangle/2$ as a function of $\sqrt{s_{\mathrm{NN}}}$ follows the trend established in previous measurements in heavy-ion collisions. 
Theoretical models based on mechanisms for particle production in nuclear collisions that involve the formation of quark--gluon plasma medium and models based on individual nucleon--nucleon interactions are compared to the data.
\end{abstract}
\end{titlepage}

\setcounter{page}{2} 


\section{Introduction}
\label{intro}
The multiplicity of produced particles is a primary observable used to characterise the system created in a high-energy heavy-ion collision and its geometry.
This quantity constrains the particle production mechanisms and is employed to estimate the initial energy density of the system~\cite{ALICE:2022wpn,Bjorken:1982qr}. 

The ALICE apparatus has been upgraded during the Long Shutdown 2 (LS2) of the Large Hadron Collider (LHC)~\cite{ALICE:2012dtf,ALICE:2023udb}. 
Key upgrades include a new tracking system in the central barrel~\cite{ALICE:2013nwm,Lippmann:2014lay}, new forward
trigger detector~\cite{Trzaska:2017reu}, continuous readout~\cite{Antonioli:2013ppp}, and a novel online-offline software framework (O\textsuperscript{2})~\cite{ALICE:o2} to handle both data reconstruction and analysis.
The upgraded ALICE detector resumed data-taking in 2022 with proton--proton (pp) collisions at a centre-of-mass energy $\sqrt{s}=$ 0.9 and 13.6 TeV and lead--lead (Pb--Pb) collisions at a centre-of-mass energy per nucleon pair $\sqrt{s_{\mathrm{NN}}}=5.36$ TeV.

This paper presents not only the first measurements by ALICE in Pb--Pb collisions at the highest energy at the LHC but also provides important validation of the upgraded apparatus and software. 
In detail, the pseudorapidity density of primary charged particles is presented as a function of pseudorapidity, exploiting also their dependence on the overlapping area of the nuclei, defined by centrality.
In accordance with the ALICE definition, primary particles are defined as charged particles with a mean proper lifetime $\tau$ larger than 1\,cm/$c$, which are either a) produced directly in the collision, or b) from decays of particles with $\tau$ smaller than 1\,cm/$c$, excluding particles produced in interactions with material~\cite{ALICE-PUBLIC-2017-005}.
Moreover, charged-particle pseudorapidity density is studied as a function of the number of participating nucleons in the collision, as determined using the Glauber model~\cite{Miller:2007ri,dEnterria:2020dwq,Loizides:2017ack}.
Where available, results are compared to those at the same collision energy by the CMS experiment~\cite{CMS:2024ykx}, or to results from different energies and colliding systems by experiments at LHC ~\cite{ALICE:2015juo, ALICE:2010khr, ALICE:2010mlf, ALICE:2022kol, ALICE:2015olq, ATLAS:2011ag, CMS:2011aqh} and at Relativistic Heavy Ion Collider (RHIC)~\cite{BRAHMS:2001gci, BRAHMS:2001llo, PHENIX:2000owy, PHOBOS:2010eyu, STAR:2008med}, as well as earlier experiments at CERN~\cite{NA50:2002edr}. 
To interpret the underlying particle production mechanisms, results are compared to models that assume quark--gluon plasma (QGP) formation with hydrodynamic evolution of the system, and models that do not incorporate QGP. 

The paper is organised as follows: Sec.~\ref{exp} presents the experimental setup focusing on the upgraded detectors, data samples and the used analysis method.
Section~\ref{analysis} describes the correction procedure in detail, while systematic uncertainties are discussed in Sec.~\ref{systematics}. 
Finally, results are presented in Sec.~\ref{results}, and the conclusions are summarised in Sec.~\ref{conclusion}.

\section{Experimental setup}
\label{exp}
\subsection{ALICE detectors in LHC Run 3}
The ALICE detector~\cite{ALICE:2012dtf,ALICE:2023udb} consists of a central barrel contained in a solenoidal magnet ($B=0.5$~T) and a forward muon system with a dipole magnet providing a total bending power of 3~{T\,m}. 
During LS2 the central barrel detectors have been upgraded and designed for efficient tracking in the high track-density environment of heavy-ion collisions, covering transverse momenta (\pt) from $\sim 100$~{MeV/$c$} to $\sim 100$~{GeV/$c$} with excellent hadron and electron identification capabilities.

The new Inner Tracking System (ITS2) consists of seven layers of ALPIDE monolithic active pixel sensors (MAPS)~\cite{ALICE:2013nwm}.
The ITS2 has improved pointing resolution owing to its reduced distance to the interaction point and better primary vertex position resolution with respect to that of the ITS used in the LHC Runs 1 and 2. 
It can also handle the hit densities produced from Pb--Pb collisions at rates up to 50~kHz in nominal operating conditions.

In the radial direction, the ITS2 is followed by the Time Projection Chamber (TPC) extending from 0.85\,m to 2.5\,m in radius over a length of 5\,m~\cite{Lippmann:2014lay,Riegler:2015lay}.
The TPC was upgraded from multiwire proportional chambers to readout chambers based on Gas Electron Multiplier (GEM) foils that reduce the ion backflow and resulting space charge in the drift volume to a correctable level for nominal operating conditions~\cite{ALICETPC:2020ann}.

A set of new forward detectors form a Fast Interaction Trigger (FIT) system~\cite{Trzaska:2017reu}, which provides precise timing for continuous readout and is also used for centrality estimation, event selection, and collision time determination. This analysis uses the FIT time-zero (FT0) detector which consists of two arrays of fast Cherenkov counters featuring unique timing capabilities: FT0A, covering the pseudorapidity interval $3.5<\eta<4.9$, and FT0C, covering $-3.3<\eta<-2.1$.

Furthermore, the readout infrastructure was completely renewed to support the continuous readout of the core detectors~\cite{Antonioli:2013ppp}. 
In nominal operating conditions, each TPC drift time period of $\approx 83$~\text{$\mu$s} contains, on average, five Pb--Pb events. 
Thus, a continuous, non-triggered readout strategy was implemented, combined with online data compression for the upgraded readout and data acquisition system.
In order to synchronise the continuous data stream across all readout and processing branches, the data stream is divided into time frames of a nominal length of 32 LHC orbits ($\approx2.8$\,ms)~\cite{Ronchetti:2024gak}.

\subsection{Data sample and analysis method}
\label{datasample}
This analysis uses a sub-sample of data from Pb--Pb collisions at $\sqrt{s_{\mathrm{NN}}}=5.36$ TeV recorded by ALICE in 2023. The analysed data sample consists of $1.1\times10^9$ minimum bias Pb--Pb events, selected from runs with interaction rates ranging from 12 to 43 kHz. This corresponds to an integrated luminosity ($L_{\mathrm{int}}$) of about 130 $\mu$b$^{-1}$, estimated before applying the event selection criteria.

The transition to continuous readout necessitates changes in the reconstruction strategy~\cite{Alkin:2021mfo}. 
In the continuous readout~\cite{Antonioli:2013ppp}, the association of tracks (reconstructed trajectories of charged particles) to the collision vertex relies on track timestamps in relation to FIT vertex signals and, thus, is affected by the time resolution.
Previously, in the triggered readout, this association was unambiguous, with the sole exception of same-bunch pileup.
As a result, now the full time frame, containing several collisions, needs to be reconstructed together.
By construction, this introduces a track-to-vertex association ambiguity. 
Depending on the interaction rate, a single track with large enough time uncertainty may be compatible with several recorded collision vertices. 
The fraction of tracks that have a better compatibility with a vertex different from the most probable one, based on the distance of the closest approach (DCA) criteria, is at sub-percent level for the 2023 data. 
Therefore, track ambiguity is neglected for the tracks that are used in this analysis. 

A set of requirements is used to select the minimum bias events.
The vertex trigger is produced by FIT when
both FT0A and FT0C arrays detect at least one hit in the nominal collision time window, synchronised with the LHC clock. The time difference between the two signals defines
the position of the collision vertex.
Collisions are removed based on their proximity in time to the ITS readout frame border and the time frame border. 
This avoids partially recorded events. 
Events with a longitudinal position within 10\,cm of the nominal interaction point are selected, based on the typical width of the vertex position Gaussian distribution of $\sim5$\,cm.
Additionally, the events where the vertex has no contributing global tracks (i.e. tracks that combine hits in ITS and TPC) and events where the vertex position estimate from FIT is more than 1\,cm away from the tracked vertex position are ignored due to lower overall quality. 

The effect of the detector occupancy on the tracking efficiency is also considered, as a consequence of the continuous readout approach combined with the high interaction rate.
The occupancy effect is similar to that of event pile-up, as the collision under consideration may share the detector readout window with collisions with very high multiplicities, resulting in a decrease in tracking capability in both ITS and TPC.
An occupancy of less than 500 tracks has been chosen in order to maintain approximately uniform tracking efficiency across the full sample. 

After applying the event selection criteria described above, the data sample consists of about 0.2 $\times$ 10$^9$ events. The selected event sample is then classified into centrality classes using the raw amplitudes of signals in the FT0C detector, which are proportional to the number of charged particles detected by the FT0C.
Following the method developed previously in Runs 1 and 2 measurements~\cite{ALICE:2013hur,ALICE-PUBLIC-2018-011}, the measured FT0C amplitude is fitted with the Glauber model~\cite{Loizides:2017ack} coupled with a two-component model assuming that the effective number of particle-producing sources ($N_{\mathrm{sources}}$) is given by
\begin{equation}
N_{\mathrm{sources}} = f \times N_{\mathrm{part}} + (1-f) \times N_{\mathrm{coll}},
\label{nsource_eqn}
\end{equation}
where $N_{\mathrm{part}}$ is the number of participating nucleons, $N_{\mathrm{coll}}$ is the number of binary nucleon--nucleon collisions, and $f$ constrains the relative contributions of $N_{\mathrm{part}}$ and $N_{\mathrm{coll}}$. The number of particles produced by each source is parameterised by the negative binomial distribution (NBD) with parameters $\mu$ and $k$, where $\mu$ is the mean multiplicity per source and $k$ controls the contribution at high multiplicity. 

In the Glauber model calculation, the nuclear density for $^{208}$Pb is modeled by modified Woods-Saxon or 2-parameter Fermi distribution (2pF)~\cite{Loizides:2017ack}
\begin{equation}
\rho (r) =  \rho_0 \frac{1+w(r/R)^2}{1+\mathrm{exp}(\frac{r-R}{a})}.
\label{nucl_density_eqn}
\end{equation}

The parameter $\rho_0$ is the nucleon density, which provides the overall normalisation.
The nuclear radius $R$ and skin thickness $a$ are 6.62 $\pm$ 0.06 fm and 0.546 $\pm$ 0.010 fm, respectively, based on data from low-energy electron--nucleus scattering experiments~\cite{DeVries:1987atn}. The additional parameter $w$ is needed to describe nuclei whose maximum density is reached at radii $r>0$ ($w=0$ for Pb).
For Pb--Pb collisions at \linebreak $\sqrt{s_{\mathrm{NN}}}=5.36$ TeV, we use an inelastic nucleon--nucleon cross section ($\sigma_{\mathrm{INEL}}$) of 68.2 $\pm$ 0.6 mb~\cite{Loizides:2017ack}, estimated by interpolation of pp data at different centre-of-mass energies and from cosmic rays, and subtracting the elastic scattering cross section from the total cross section. 

The NBD-Glauber fit provides a good description of the observed FT0C amplitude in the region corresponding to the top 90\% of the hadronic cross section. The average $\langle N_{\mathrm{part}} \rangle$ in a given centrality class is obtained from the Glauber model imposing the same cuts applied to the data on the simulated FT0C response.

Tracking in the pseudorapidity range of this study is provided by two sub-detectors, ITS and TPC.
For a vertex at the nominal interaction point, the TPC geometrical acceptance is \mbox{$\left|\eta\right| < 0.9$} while that of the ITS is \mbox{$\left|\eta\right| < 1.22$}.
The tracks available for analysis can be divided into three exclusive classes: global tracks, built from ITS and TPC contributions, ITS-only tracks and TPC-only tracks.
In this analysis, TPC-only tracks are not used because their pointing resolution is low, as the innermost TPC radius is quite far from the collision vertex, and they cannot be reliably assigned to a collision.
The selected track sample contains global tracks passing quality cuts for ITS and TPC contributions, and ITS-only tracks passing quality cuts for ITS contributions. The quality cuts for TPC contributions include at least 70 pad hit clusters in the TPC with a good track quality ($\chi^2$/dof $<$ 4). The quality cuts for ITS contributions include at least one hit among the three innermost ITS layers and $\chi^2$/dof $<$ 36.
Additionally, both categories of tracks must satisfy a set of conditions on the DCA to the primary vertex in order to ensure that tracks are originating from the primary interaction.
The DCA is required to be within 0.2\,cm in the longitudinal direction and lower than 0.0105 + 0.035$\pt^{-1.1}$ in the transverse direction with \pt in GeV/$c$. No additional selection on the \pt of reconstructed
tracks is performed in order to fully leverage the acceptance of the ALICE detector at
low momenta.

\section{Correction procedure}
\label{analysis}
The pseudorapidity density of the reconstructed tracks ($\mathrm{d}N_{\mathrm{ch}}^{\mathrm{uncorrected}}/\mathrm{d}\eta$) is corrected for detector effects and normalised by the number of events in a corresponding centrality interval.
Detector effects are estimated using simulations with \angantyr~\cite{Bierlich:2018xfw} in which the produced particles are transported through the geometric detector model using GEANT~4~\cite{Allison:2016lfl}, digitised and reconstructed to obtain detector response. The simulated detector response takes account of known conditions at the time of the data-taking, including inefficiencies, acceptance, electronic noise, and other smearing effects.

Three types of correction are applied: (a) track-to-particle correction, which accounts for detector and reconstruction inefficiencies and background from secondary particles (i.e. particles produced in interactions with the detector material); (b) contamination by weak-decay products of long-lived particles, primarily \kzero, \lmb, and \almb; (c) correction for event and signal loss due to the applied event selection criteria.
The track-to-particle correction factor is based on ratios of \dndeta values obtained at the particle (generated) level and detector (reconstructed) level as a function of pseudorapidity. One can assume that $O^{\mathrm{gen}}_{\mathrm{mc}}$ be the distribution of $\mathrm{d}N_{\mathrm{ch}}^{\mathrm{gen}}/\mathrm{d}\eta$ obtained by the \angantyr at the generated level and $O^{\mathrm{reco}}_{\mathrm{mc}}$ be the distribution of $\mathrm{d}N_{\mathrm{ch}}^{\mathrm{reco}}/\mathrm{d}\eta$ obtained at the reconstructed level. The corrected \dndeta is obtained by multiplying the $\mathrm{d}N_{\mathrm{ch}}^{\mathrm{uncorrected}}/\mathrm{d}\eta$ with the factor $\frac{O^{\mathrm{gen}}_{\mathrm{mc}}}{O^{\mathrm{reco}}_{\mathrm{mc}}}$. The correction factor for 0--5\% (80--90\%) Pb--Pb collisions is of the order of 1.05 (1.1) and is almost $\eta$-independent.

Based on the definition of primary particles, the long-lived ($\tau>1$\,cm/$c$) decay products of \kzero, \lmb, and \almb are not considered as primary. 
To account for this, we need to subtract the number of strange particle decay tracks from both $\mathrm{d}N_{\mathrm{ch}}^{\mathrm{uncorrected}}/\mathrm{d}\eta$ and $O^{\mathrm{reco}}_{\mathrm{mc}}$. 
This contamination is estimated in simulations by obtaining the distribution of reconstructed tracks ($O^{\mathrm{strange}}_{\mathrm{mc}}$) identified at the generator level as originating from strange decays.
As the yield of strange particles from \angantyr does not match the observed values in data, an additional correction factor, the strangeness enhancement factor (SEF), is determined. This is obtained by comparing raw yields of strange particles in data and
simulation and is determined to be 1.1. The weak-decay contamination in data is, therefore, estimated as SEF $\times$ $O^{\mathrm{strange}}_{\mathrm{mc}}$.

The distribution of $\mathrm{d}N_{\mathrm{ch}}^{\mathrm{uncorrected}}/\mathrm{d}\eta$ is further corrected for the event and signal loss, evaluated from the MC simulation. The signal loss correction ($f_{\mathrm{SL}}$) is calculated for each centrality class by obtaining the ratio of the $\mathrm{d}N_{\mathrm{ch}}^{\mathrm{gen}}/\mathrm{d}\eta$ before trigger and event selection with the corresponding $\mathrm{d}N_{\mathrm{ch}}^{\mathrm{gen}}/\mathrm{d}\eta$ after applying all the selections. The event loss correction ($f_{\mathrm{ev}}$) corresponds to the fraction of generated events that do not pass the event selection criteria. The ratio $f_{\mathrm{ev}}/f_{\mathrm{SL}}$ is $\eta$-independent and amounts to about 0.96 in the 80--90\% centrality class, while it is negligible for other centrality classes.
The corrected \dndeta is finally obtained using Eq.~\eqref{dndeta-correct-eqn} as given below

\begin{equation}
  \frac{\mathrm{d}N_{\mathrm{ch}}}{\mathrm{d}\eta} = \biggl [\frac{\mathrm{d}N_{\mathrm{ch}}^{\mathrm{uncorrected}}}{\mathrm{d}\eta} - (\mathrm{SEF} \times O^{\mathrm{strange}}_{\mathrm{mc}})\biggr ] \times \frac{O^{\mathrm{gen}}_{\mathrm{mc}}}{O^{\mathrm{reco}}_{\mathrm{mc}} - O^{\mathrm{strange}}_{\mathrm{mc}}} \times \frac{f_{\mathrm{ev}}}{f_{\mathrm{SL}}}.
  \label{dndeta-correct-eqn}
\end{equation}

\section{Systematic uncertainties}
\label{systematics}
The contributions of the different sources of systematic uncertainties associated with the measurements of charged-particle pseudorapidity density are summarised in Table~\ref{table:sysunc}.
A study is performed based on occupancy of more than 500 tracks other than the nominal one as described in Sec.~\ref{datasample}, and we observe that the results are consistent with each other. 
Therefore, the occupancy effect on the final measurements is not added to the systematic uncertainty.
The details about the different conditions varied in the analysis to estimate the systematic uncertainties are described below.

\begin{itemize}
    \item \textit{Acceptance and efficiency}:
    This uncertainty arises from the limited detector efficiency and non-uniformity in azimuthal acceptance of ITS and TPC. 
    It is estimated by measuring the \dndeta in six different azimuthal sectors and comparing them to the measurement in the full azimuth. This uncertainty is considered as correlated over both $\eta$ and centrality.
    \item \textit{$z_{\mathrm{vtx}}$ dependence}:
    The effect of detector acceptance is also studied by carrying out the analysis for different slices of the $z$ position of the interaction vertex distribution and comparing the results with the one obtained with nominal $z_{\mathrm{vtx}}$ cut ($|z_{\mathrm{vtx}}|<10$\,cm). This uncertainty is uncorrelated over both $\eta$ and centrality.
    \item \textit{\pt extrapolation}:
    Below 100 MeV/$c$, the charged-particle reconstruction efficiency sharply drops because of the bending of charged-particle trajectories under the effect of the magnetic field and, to a lesser extent, due to the scattering and absorption in the detector material. 
    To study the effect of the undetected particles at low \pt, the production of primary charged particles with \pt~$<$~100~MeV/$c$ is varied in the event generator by a conservative amount (+100\% and $-$50\%), which is adopted from the previous Run~2 measurements~\cite{ALICE:2015olq}. This uncertainty is correlated over both $\eta$ and centrality.
    \item \textit{Strangeness content}:
    The uncertainty associated with the correction for the contamination from weak decays of strange hadrons is determined by varying the strange particle content in the event generator. 
    The yield of strange particles in data is found to be 10\% larger than in simulation, and this discrepancy is varied by $\pm$20\% to estimate the uncertainty. 
    This uncertainty is considered correlated over both $\eta$ and centrality.
    \item \textit{Particle-species composition}:
    The composition of the produced particle species affects the determination of efficiency because different particle species have different efficiencies that depend on the applied effective \pt thresholds and the decay kinematics. 
    The influence of this uncertainty is estimated by changing the relative abundances of charged kaons, protons and other charged particles with respect to charged pions by $\pm$ 30\% in the simulation. This uncertainty is correlated over both $\eta$ and centrality.
    \item \textit{Track selection}: 
    The systematic uncertainty related to the track selection criteria is determined by varying the track matching requirements in the ITS detector. 
    In particular, the nominal analysis requires at least one hit in the first three ITS layers. 
    Two alternative criteria are tested: (a) requiring at least one hit among all the ITS layers, and (b) requiring one hit on all the ITS layers. This uncertainty is uncorrelated over both $\eta$ and centrality.
    \item \textit{Centrality determination}:
    The centrality determination introduces an uncertainty via the fitting of the FT0C amplitude distribution to the hadronic cross section due to the contamination from electromagnetically induced reactions at small multiplicities.
    The uncertainties associated with centrality determination and $\langle N_{\mathrm{part}} \rangle$ calculation are evaluated by varying certain parts of the Glauber fitting procedure. In particular, in the default FT0C estimator, the signal region in which data and the NBD-Glauber fit are matched is up to the top 90\% of the hadronic cross section, while in the systematic variations, it is evaluated on the full centrality range. A second variation is done to allow the parameter $k$ of the NBD to be free, instead of fixed to 1.5 as in the default FT0C estimator.  Finally, $\sigma_{\mathrm{INEL}}$ and $a$, $R$ in Eq.~\eqref{nucl_density_eqn} are varied within their systematic uncertainty values. These variations result in an uncertainty on \dndeta of 0.1\% for the most central collisions, increasing towards the peripheral collisions and reaching 11.5\% for the 80--90\% sample. This uncertainty is correlated over $\eta$ but uncorrelated over centrality.
  \item \textit{Estimation of event and signal loss corrections}:
    These corrections depend on how the generator level centrality in MC is estimated. The event and signal loss corrections are obtained using two independent proxies for generator-level centrality: the impact parameter distribution and the charged-particle multiplicity distribution within the FT0C acceptance. The uncertainty associated with this estimation is then evaluated from the difference between the correction factors obtained using these two methods. This difference results in an uncertainty on \dndeta of 3.5\% for the most peripheral collisions, while being negligible for the other centrality classes. This uncertainty is correlated over $\eta$ but uncorrelated over centrality.
\end{itemize}

The contributions of different sources of correlated and uncorrelated uncertainties are summed in quadrature to obtain the total systematic uncertainty in each centrality interval, which varies from 2.6\% in the most central collisions to 12.4\% in the most peripheral collisions.

\begin{table}[h!]
\scriptsize
\centering
\caption{Contributions to systematic uncertainties (in percent) in the measurements of charged-particle pseudorapidity  density for different centrality classes in Pb--Pb collisions at $\sqrt{s_{\mathrm{NN}}}=5.36$ TeV.}
\label{table:sysunc}
\renewcommand{\arraystretch}{1.4}
\begin{tabular}{lcccccccccc}
\hline
\hline
\textbf{Sources} & \multicolumn{10}{c}{Systematic uncertainty at $|\eta|<0.5$ (in \%)} \\
\cline{2-11} \\[-0.4cm]
& 0--5\% & 5--10\% & 10--20\% & 20--30\% & 30--40\% & 40--50\% & 50--60\% & 60--70\% & 70--80\% & 80--90\% \\
\hline
Acceptance and efficiency & 2.2 & 2.1 & 2.0 & 2.0 & 1.9 & 1.9 & 2.0 & 1.9 & 1.9 & 2.0 \\
$z_{\mathrm{vtx}}$ dependence & $<0.1$ & $<0.1$ & $<0.1$ & $<0.1$ & $<0.1$ & $<0.1$ & 0.1 & 0.1 & 0.1 & 0.2 \\
\pt extrapolation & 1.2 & 1.2 & 1.2 & 1.2 & 1.2 & 1.3 & 1.3 & 1.3 & 1.3 & 1.3 \\
Strangeness content & 0.4 & 0.4 & 0.3 & 0.3 & 0.3 & 0.3 & 0.2 & 0.2 & 0.2 & 0.2 \\
Particle-species composition & 0.4 & 0.4 & 0.4 & 0.4 & 0.4 & 0.4 & 0.4 & 0.5 & 0.5 & 0.5 \\
Track selection & 0.6 & 0.3 & 0.7 & 1.1 & 1.3 & 1.5 & 1.6 & 1.8 & 1.9 & 2.0 \\
Event and signal loss corrections & $<0.1$ & $<0.1$ & $<0.1$ & $<0.1$ & $<0.1$ & $<0.1$ & $<0.1$ & 0.1 & 0.1 & 3.5 \\
Centrality determination & 0.1 & 0.4 & 0.8 & 1.3 & 2.1 & 3.1 & 4.4 & 6.3 & 8.7 & 11.5 \\
\hline
Total & 2.6 & 2.5 & 2.6 & 2.9 & 3.4 & 4.1 & 5.2 & 6.9 & 9.2 & 12.4 \\
\hline
\hline
\end{tabular}
\end{table}

\section{Results and discussion}
\label{results}
\begin{figure}[h!]
  \centering
  \includegraphics[scale=0.5]{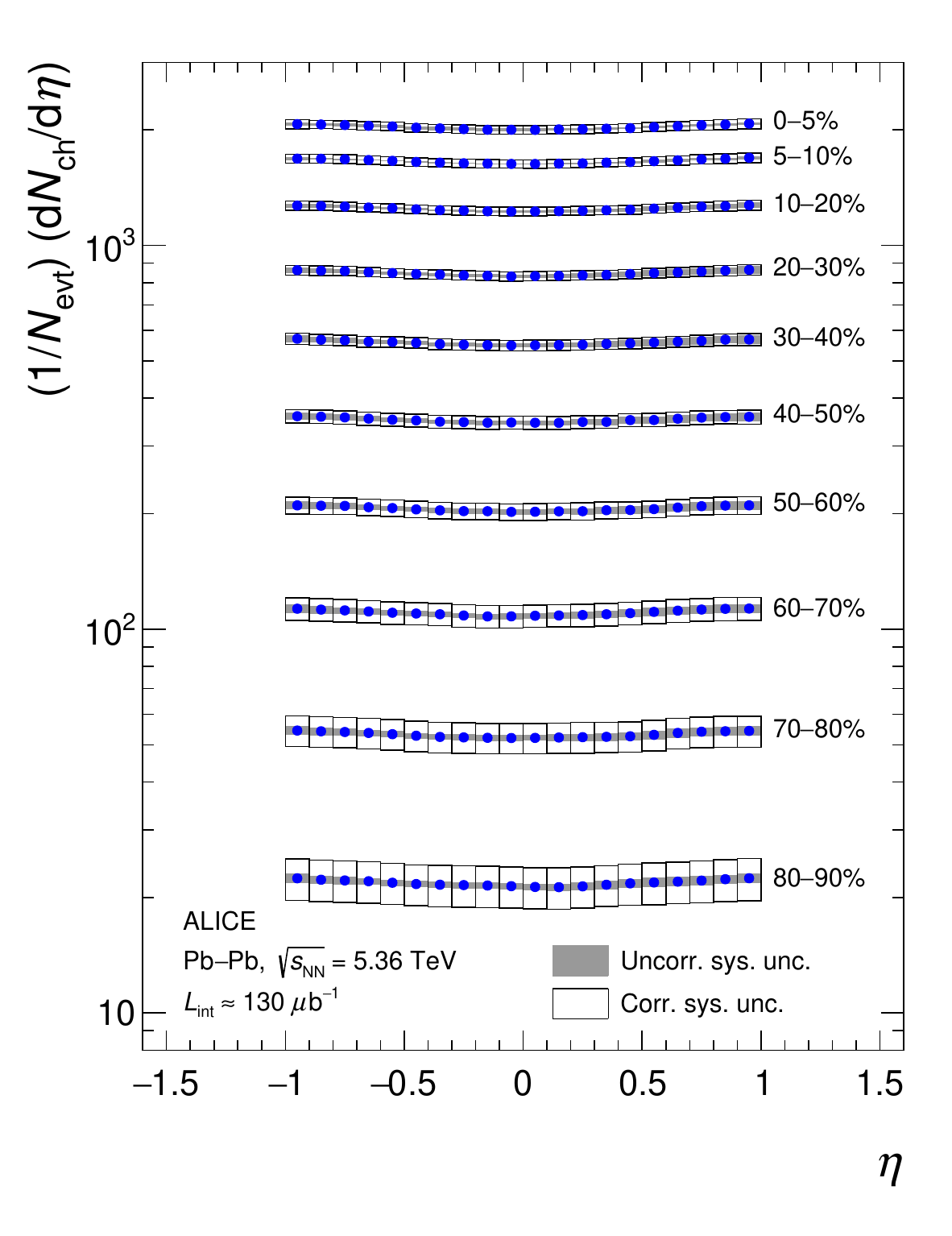}
  \caption{Charged-particle pseudorapidity density at midrapidity for ten centrality classes in Pb--Pb collisions at $\sqrt{s_{\mathrm{NN}}}=5.36$ TeV. Grey and open boxes show the uncorrelated and correlated systematic uncertainties over pseudorapidity, respectively. Statistical uncertainties are smaller than the marker size.}
  \label{dnch_deta_dist}
\end{figure}

Figure~\ref{dnch_deta_dist} presents the measured primary charged-particle pseudorapidity density, \dndeta, as a function of pseudorapidity for ten centrality classes in Pb--Pb collisions at $\sqrt{s_{\mathrm{NN}}}=5.36$ TeV.
The \dndeta exhibits a rapid increase with collision centrality for all $\eta$. 
The conventional approach for characterising the particle production in heavy-ion collisions is to measure the average charged-particle pseudorapidity density at midrapidity, which is defined as \dndeta averaged over $|\eta|<0.5$ and denoted by \avdndeta. 
The measured \avdndeta are 2010 $\pm$ 53 (syst) and 21.5 $\pm$ 2.7 (syst) for the 0--5\% and 80--90\% centrality classes, respectively. 
The statistical uncertainties are negligible. 
Table~\ref{avg-correct-dndetabynpart} provides the results of \avdndeta for all centrality intervals considered in this analysis.

\begin{table}[h!]
 \caption{The \avdndeta and (2/$\langle N_{\mathrm{part}} \rangle$)\avdndeta values measured in $|\eta|<0.5$ for ten centrality classes in \PbPb collisions at $\sqrt{s_{\mathrm{NN}}}=5.36$ TeV. The values of $\langle N_{\mathrm{part}} \rangle$ obtained with the Glauber model are also given. The uncertainties are total systematic uncertainties, and statistical uncertainties are negligible.}
 \label{avg-correct-dndetabynpart}
  \begin{center}
    \renewcommand{\arraystretch}{1.2}
    \begin{tabular}{|c|c|c|c|}
      \hline
      Centrality & \avdndeta & $\langle N_{\mathrm{part}} \rangle$ & (2/$\langle N_{\mathrm{part}} \rangle$)\avdndeta \\
      \hline
      0--5\% & 2010 $\pm$ 53 & 383.6 $\pm$ 0.8 & 10.5 $\pm$ 0.3 \\
      5--10\% & 1638 $\pm$ 42 & 332.4 $\pm$ 1.7 & 9.9 $\pm$ 0.3 \\
      10--20\% & 1232 $\pm$ 33 & 263.1 $\pm$ 1.3 & 9.4 $\pm$ 0.3 \\
      20--30\% & 836 $\pm$ 25 & 188.4 $\pm$ 1.3 & 8.9 $\pm$ 0.3 \\
      30--40\% & 552 $\pm$ 19 & 130.6 $\pm$ 1.4 & 8.4 $\pm$ 0.3 \\
      40--50\% & 347 $\pm$ 14 & 86.5 $\pm$ 1.5 & 8.0 $\pm$ 0.4 \\
      50--60\% & 203 $\pm$ 11 & 53.7 $\pm$ 1.1 & 7.6 $\pm$ 0.4 \\
      60--70\% & 109 $\pm$ 8 & 30.5 $\pm$ 0.8 & 7.1 $\pm$ 0.5 \\
      70--80\% & 52 $\pm$ 5 & 15.4 $\pm$ 0.4 & 6.8 $\pm$ 0.7 \\
      80--90\% & 21.5 $\pm$ 2.7 & 6.8 $\pm$ 0.2 & 6.3 $\pm$ 0.8 \\
      \hline
    \end{tabular}
  \end{center}
\end{table}

\begin{figure}[h!]
  \centering
  \includegraphics[scale=0.5]{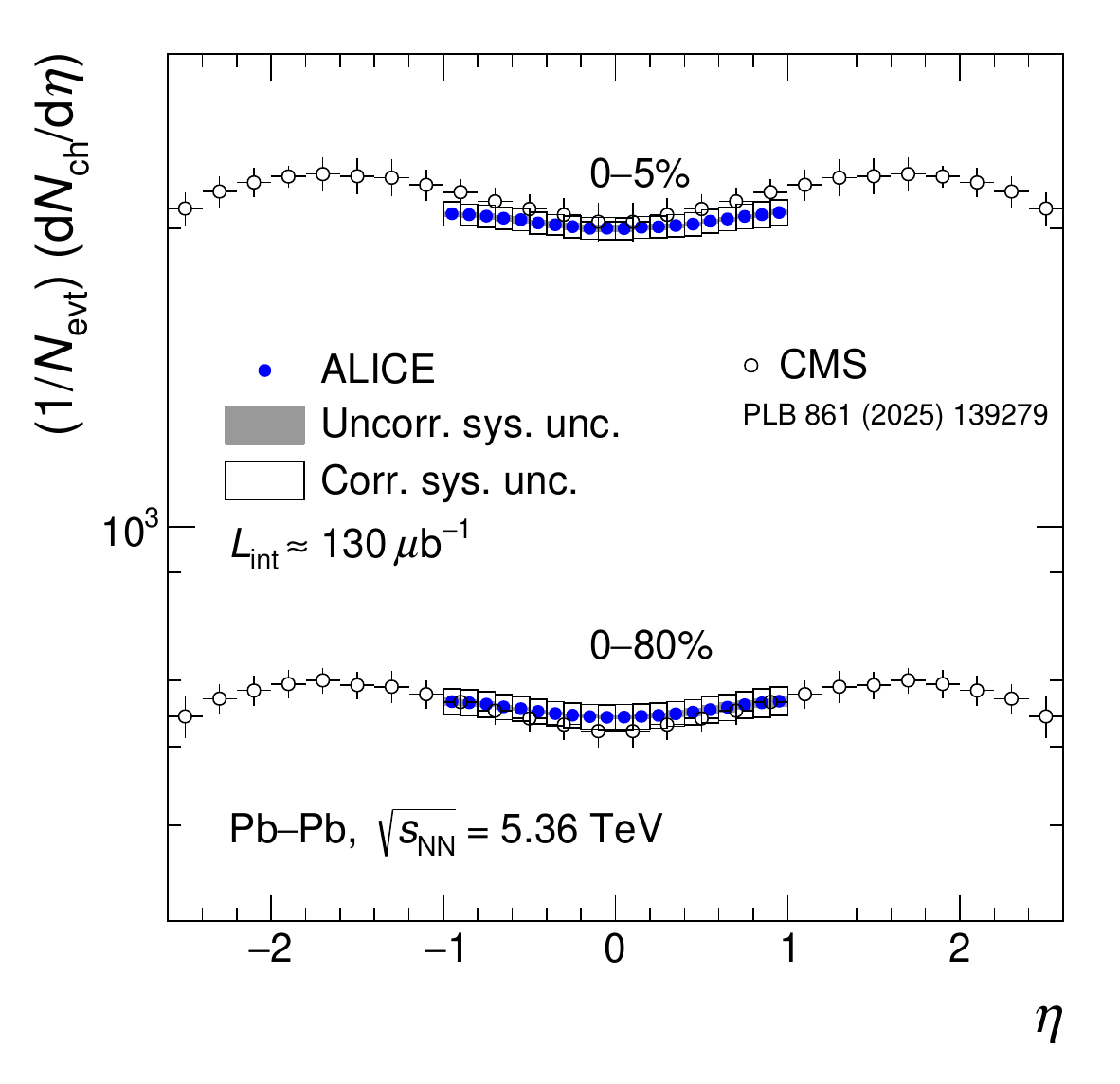}
  \caption{Comparison of charged-particle pseudorapidity density measured in 0--80\% and 0--5\% Pb--Pb collisions over $|\eta|<1$ with similar measurements from the CMS experiment in $|\eta|<2.6$. For ALICE data, grey and open boxes show the uncorrelated and correlated systematic uncertainties over pseudorapidity, respectively. Statistical uncertainties are smaller than the marker size. For CMS data, the vertical line denotes the total systematic uncertainties.}
  \label{dnch_deta_with_cms}
\end{figure}

Figure~\ref{dnch_deta_with_cms} compares the measurements by ALICE for 0--80\% and 0--5\% central \PbPb collisions with recent results from the CMS experiment~\cite{CMS:2024ykx} where centrality classes are defined based on the transverse energy in forward calorimeters unlike the multiplicity-based selection used in this analysis. 
The CMS results cover the pseudorapidity range $|\eta|<2.6$ while the ALICE measurement is shown for $|\eta|<1$. The results from the two experiments are consistent with each other within uncertainties.

\begin{figure}[h!]
  \centering
  \includegraphics[scale=0.5]{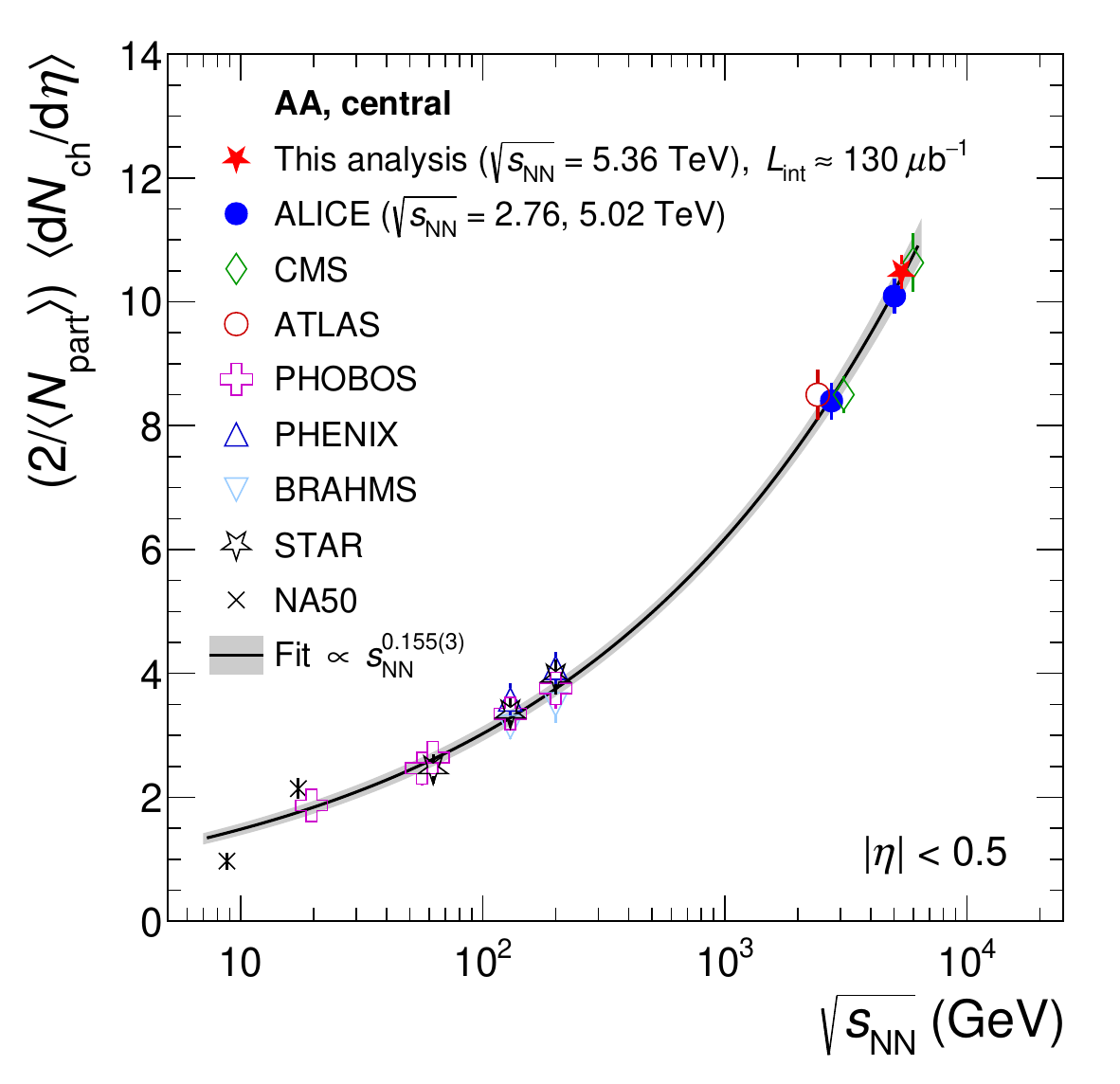}
  \caption{Values of (2/$\langle N_{\mathrm{part}} \rangle$)\avdndeta for central Pb--Pb and Au--Au collisions as a function of $\sqrt{s_{\mathrm{NN}}}$.  The $\sqrt{s_{\mathrm{NN}}}$ dependence of the heavy-ion collision data is well described by the function $s^{0.155}_{\mathrm{NN}}$ (solid line). The shaded band represents the uncertainty on the extracted power-law dependency. The central Pb--Pb measurements from ATLAS at $\sqrt{s_{\mathrm{NN}}}=2.76$ TeV and from CMS at $\sqrt{s_{\mathrm{NN}}}=2.76$ and 5.36 TeV have been shifted horizontally for clarity.}
  \label{dnch_deta_with_energy}
\end{figure}

To compare the bulk particle production in different collision systems and at different energies, the \avdndeta is normalised by the average number of participating nucleon pairs, $\langle N_{\mathrm{part}} \rangle$/2.
The $\langle N_{\mathrm{part}} \rangle$ values are listed in Table~\ref{avg-correct-dndetabynpart}.
For the most central (0--5\%) collisions, the normalised multiplicity, \linebreak (2/$\langle N_{\mathrm{part}} \rangle$)\avdndeta, is measured to be 10.5 $\pm$ 0.3 (syst) with negligible statistical uncertainty. 
This value is (4.0 $\pm$ 3.9)\% higher than that measured earlier at $\sqrt{s_{\mathrm{NN}}}=5.02$ TeV by ALICE~\cite{ALICE:2015juo}, considering fully uncorrelated uncertainties. 
In Fig.~\ref{dnch_deta_with_energy}, this value (solid star symbol) is compared to the central Au--Au and Pb--Pb collisions at lower centre-of-mass energies per nucleon pair from experiments at the RHIC~\cite{BRAHMS:2001gci,BRAHMS:2001llo,PHENIX:2000owy,PHOBOS:2010eyu,STAR:2008med}, SPS~\cite{NA50:2002edr}, and LHC~\cite{ALICE:2010khr,ALICE:2010mlf,ALICE:2015juo,ATLAS:2011ag,CMS:2011aqh}.
The dependence of (2/$\langle N_{\mathrm{part}} \rangle$)\avdndeta on $\sqrt{s_{\mathrm{NN}}}$ is fitted with a power-law ($\alpha \cdot s_{\mathrm{NN}}^{\beta}$) that gives $\beta=0.155\pm0.003$.
The uncertainty on $\beta$ is obtained, assuming that the data uncertainties are uncorrelated at different centre-of-mass energies.
This $\beta$ value can be compared to $\beta=0.102\pm0.003$ obtained for inelastic pp collisions using the same parametrisation~\cite{ALICE:2022kol}. 
This shows that the charged-particle multiplicity increases faster with energy in central heavy-ion collisions compared to pp collisions. 
The result at $\sqrt{s_{\mathrm{NN}}}=5.36$ TeV confirms the trend established by lower-energy data since $\beta$ does not change when the new measurement is included in the fit. In Run 2, the $\beta$ value was $0.155\pm0.004$~\cite{ALICE:2015juo}.

\begin{figure}[h!]
  \centering
  \includegraphics[scale=0.5]{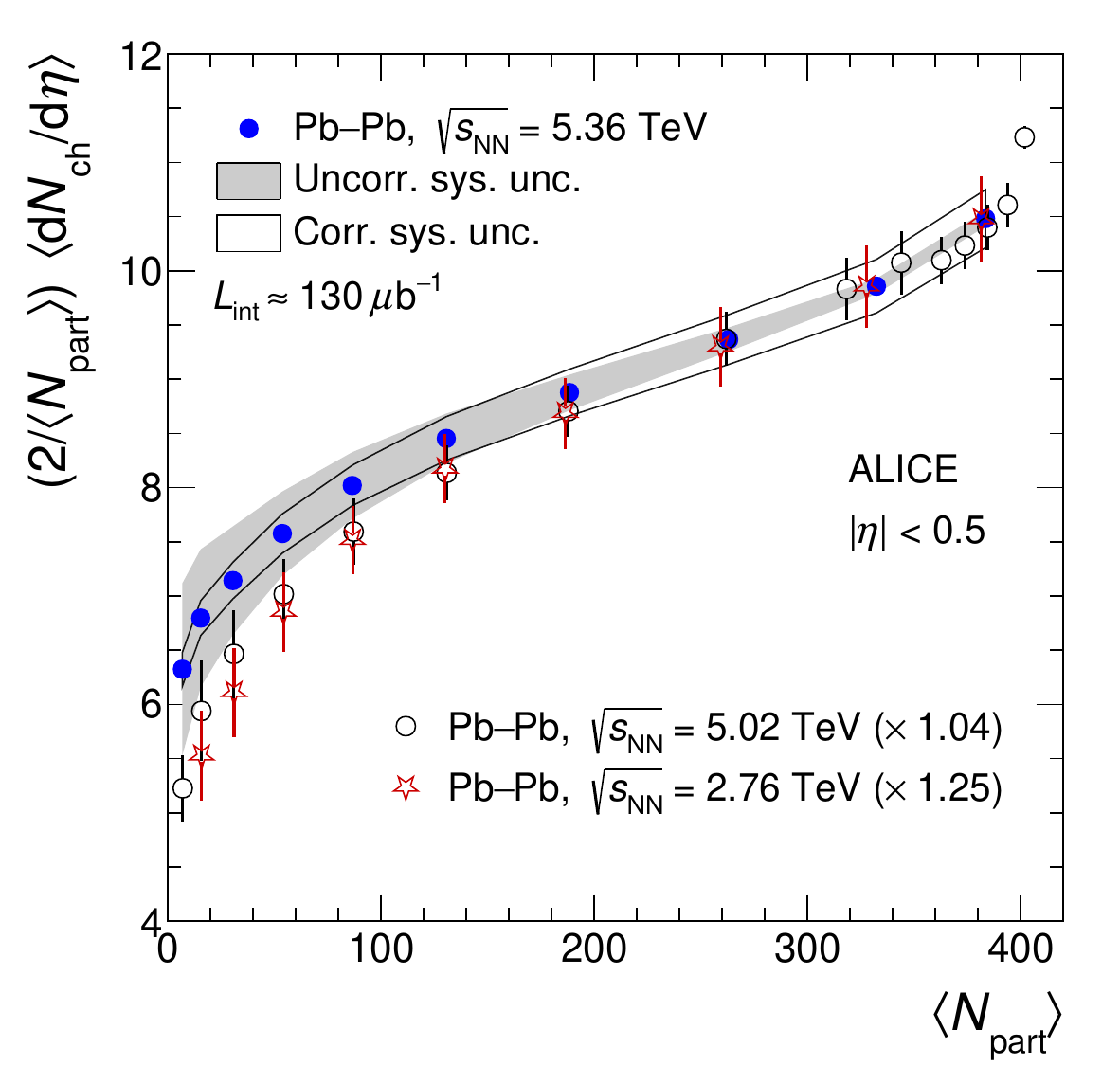}
  \caption{Dependence of (2/$\langle N_{\mathrm{part}} \rangle$)\avdndeta on the average number of participating nucleons for Pb--Pb collisions at $\sqrt{s_{\mathrm{NN}}}=5.36$ TeV. Grey bands (open boxes) show the uncorrelated (correlated) uncertainties over centrality. Statistical uncertainties are within the marker size. Data from Pb--Pb collisions at $\sqrt{s_{\mathrm{NN}}}=2.76$ TeV scaled by 1.25 and $\sqrt{s_{\mathrm{NN}}}=5.02$ TeV scaled by 1.04 are shown for comparison.}
  \label{dnch_deta_with_npart}
\end{figure}

Figure~\ref{dnch_deta_with_npart} presents the measured (2/$\langle N_{\mathrm{part}} \rangle$)\avdndeta (solid blue circles) as a function of $\langle N_{\mathrm{part}} \rangle$ in Pb--Pb collisions at $\sqrt{s_{\mathrm{NN}}}=5.36$ TeV.
A pronounced centrality dependence is observed, with (2/$\langle N_{\mathrm{part}} \rangle$)\avdndeta decreasing by a factor of $\simeq$1.75 from the most central (large $\langle N_{\mathrm{part}} \rangle$) to the most peripheral (small $\langle N_{\mathrm{part}} \rangle$) collisions.
The measurements in Pb--Pb collisions at $\sqrt{s_{\mathrm{NN}}}=2.76$ TeV~\cite{ALICE:2010mlf} scaled by 1.25 and \linebreak $\sqrt{s_{\mathrm{NN}}}=5.02$ TeV~\cite{ALICE:2015juo,ALICE:2016fbt,ALICE:2018cpu} scaled by 1.04 are also shown.
These scaling factors are derived from the observed $s^{0.155}_{\mathrm{NN}}$ dependence of the multiplicity in the most central collisions (see Fig.~\ref{dnch_deta_with_energy}).
After scaling, the trend of $\sqrt{s_{\mathrm{NN}}}=2.76$ and 5.02 TeV data are in good agreement with $\sqrt{s_{\mathrm{NN}}}=5.36$ TeV measurements within uncertainties, considering them fully uncorrelated.

\begin{figure}[h!]
  \centering
  \includegraphics[scale=0.5]{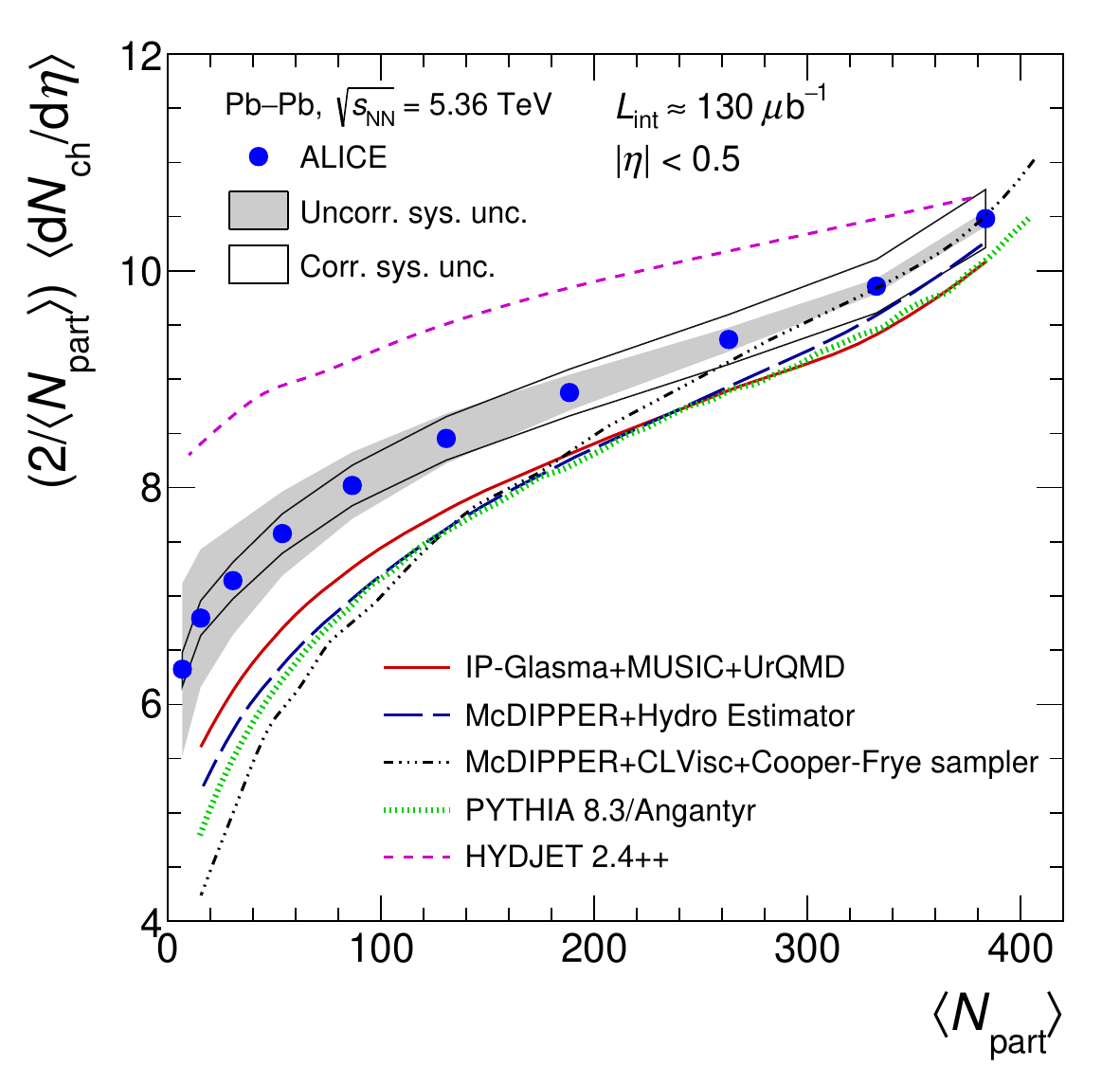}
  \caption{Comparison of (2/$\langle N_{\mathrm{part}} \rangle$)\avdndeta with model calculations for Pb--Pb collisions at $\sqrt{s_{\mathrm{NN}}}=5.36$ TeV. Uncertainties in the data are shown as in Fig.~\ref{dnch_deta_with_npart}.}
  \label{dnch_deta_with_npart_withmc}
\end{figure}

In Fig.~\ref{dnch_deta_with_npart_withmc}, the centrality dependence of (2/$\langle N_{\mathrm{part}} \rangle$)\avdndeta is compared to predictions from the HYDJET++ (v2.4)~\cite{Lokhtin:hydjet} and the \angantyr~\cite{Bierlich:2018xfw} event generators and to hydrodynamic calculations using IP-Glasma~\cite{Schenke:IPglasma1} and McDIPPER~\cite{Garcia-Montero:MCdipper1} models. 
Among these, the \angantyr and McDIPPER models are more recent developments that
were not available at the time of our previous publication at $\sqrt{s_{\mathrm{NN}}}=5.02$ TeV in 2016, allowing
us to incorporate additional theoretical perspectives in the comparison with the new data presented here.
The HYDJET++~\cite{Lokhtin:hydjet} describes a heavy-ion collision as a superposition of a soft, hydro-inspired component and a hard component arising from multi-parton fragmentation. 
It treats both these components independently. The hard component is simulated using PYthia
QUENched (PYQUEN) model~\cite{Lokhtin:2005px}, while the soft component represents a thermalised hadronic state formed on freeze-out hypersurfaces parameterised through relativistic hydrodynamics. 
An effective thermal volume approximation is employed to estimate charged-hadron multiplicities. 
In \angantyr~\cite{Bierlich:2018xfw}, the nucleon--nucleon interaction model of PYTHIA~8~\cite{Bierlich:2022pfr} is extrapolated to simulate heavy-ion collisions. 
This involves modelling individual nucleon--nucleon interactions at the parton level, which are hadronised using the Lund strings fragmentation model without incorporating any final-state collective effects.
The IP-Glasma is an initial-state model~\cite{Schenke:IPglasma1} based on the Colour Glass Condensate (CGC) effective field theory to describe the initial distribution of gluons in the colliding nuclei, incorporating gluon saturation effects and impact parameter dependence. 
In this study, a hybrid approach~\cite{Schenke:IPglasma2} is utilised, coupling IP-Glasma initial conditions with the MUSIC hydrodynamic model and the UrQMD hadronic cascade.
Similarly, the McDIPPER model~\cite{Garcia-Montero:MCdipper1} provides initial-state energy and charge deposition in heavy-ion collisions based on leading-order cross section calculations within the CGC framework. 
The \dndeta is derived from the initial energy density profile using a hydrodynamical response model (McDIPPER + Hydro Estimator) as detailed in Refs.~\cite{Garcia-Montero:MCdipper1,Giacalone:MCdipper2}.
Furthermore, a detailed phenomenological study is performed based on full 3+1D simulations (McDIPPER + CLVisc + Cooper-Frye sampler~\cite{Garcia-Montero:MCdipper2}), integrating initial-state dynamics, hydrodynamics, and hadronic cascade models to compute the \dndeta.

Figure~\ref{dnch_deta_with_npart_withmc} shows that the initial-state models and \angantyr qualitatively capture the overall trend (i.e. shape) observed in the data.  However, all these models systematically underestimate the data from central to peripheral collisions. The introduction of hadronic cascade simulation within the McDIPPER framework gives a reasonable description of central Pb--Pb data. HYDJET++, on the other hand, fails to reproduce both the shape and the magnitude of the centrality dependence of multiplicity while it agrees with the measured value within the uncertainties for the most central collisions.

\section{Conclusions}
\label{conclusion}
This letter presents the measurements of charged-particle pseudorapidity density over $|\eta|<1$ as a function of centrality in Pb--Pb collisions at $\sqrt{s_{\mathrm{NN}}}=5.36$ TeV with the ALICE detector at the LHC. 
Two upgraded detectors, ITS and TPC in continuous readout mode, are used to reconstruct charged particles at midrapidity. 

For the 5\% most central Pb--Pb collisions, the \avdndeta in $|\eta|<0.5$ is 2010 $\pm$ 53 (syst), which corresponds to a value of 10.5 $\pm$ 0.3 (syst) per participating nucleon pair. 
The latter quantity is found to increase by (4.0 $\pm$ 3.9)\%, considering fully uncorrelated uncertainties, with respect to similar measurements at $\sqrt{s_{\mathrm{NN}}}=5.02$ TeV and is in agreement with the expectations from lower energy extrapolations. 
The centrality dependence of (2/$\langle N_{\mathrm{part}} \rangle$)\avdndeta agrees well with that observed at lower energy Pb--Pb collisions. 
None of the considered theoretical models quantitatively reproduces the data in the entire centrality range reported in this paper. 
The measurements of this fundamental observable demonstrate a major milestone of the upgraded ALICE detector and its capability in analysing high-interaction-rate data with the new online-offline framework.


\newenvironment{acknowledgement}{\relax}{\relax}
\begin{acknowledgement}
\section*{Acknowledgements}
We would like to thank Oscar Garcia-Montero, Jie Zhu, S$\mathrm{\ddot{o}}$ren Schlichting, Bjoern P. Schenke, Wenbin Zhao, and Chun Shen for providing their model predictions.

The ALICE Collaboration would like to thank all its engineers and technicians for their invaluable contributions to the construction of the experiment and the CERN accelerator teams for the outstanding performance of the LHC complex.
The ALICE Collaboration gratefully acknowledges the resources and support provided by all Grid centres and the Worldwide LHC Computing Grid (WLCG) collaboration.
The ALICE Collaboration acknowledges the following funding agencies for their support in building and running the ALICE detector:
A. I. Alikhanyan National Science Laboratory (Yerevan Physics Institute) Foundation (ANSL), State Committee of Science and World Federation of Scientists (WFS), Armenia;
Austrian Academy of Sciences, Austrian Science Fund (FWF): [M 2467-N36] and Nationalstiftung f\"{u}r Forschung, Technologie und Entwicklung, Austria;
Ministry of Communications and High Technologies, National Nuclear Research Center, Azerbaijan;
Rede Nacional de Física de Altas Energias (Renafae), Financiadora de Estudos e Projetos (Finep), Funda\c{c}\~{a}o de Amparo \`{a} Pesquisa do Estado de S\~{a}o Paulo (FAPESP) and The Sao Paulo Research Foundation  (FAPESP), Brazil;
Bulgarian Ministry of Education and Science, within the National Roadmap for Research Infrastructures 2020-2027 (object CERN), Bulgaria;
Ministry of Education of China (MOEC) , Ministry of Science \& Technology of China (MSTC) and National Natural Science Foundation of China (NSFC), China;
Ministry of Science and Education and Croatian Science Foundation, Croatia;
Centro de Aplicaciones Tecnol\'{o}gicas y Desarrollo Nuclear (CEADEN), Cubaenerg\'{\i}a, Cuba;
Ministry of Education, Youth and Sports of the Czech Republic, Czech Republic;
The Danish Council for Independent Research | Natural Sciences, the VILLUM FONDEN and Danish National Research Foundation (DNRF), Denmark;
Helsinki Institute of Physics (HIP), Finland;
Commissariat \`{a} l'Energie Atomique (CEA) and Institut National de Physique Nucl\'{e}aire et de Physique des Particules (IN2P3) and Centre National de la Recherche Scientifique (CNRS), France;
Bundesministerium f\"{u}r Forschung, Technologie und Raumfahrt (BMFTR) and GSI Helmholtzzentrum f\"{u}r Schwerionenforschung GmbH, Germany;
General Secretariat for Research and Technology, Ministry of Education, Research and Religions, Greece;
National Research, Development and Innovation Office, Hungary;
Department of Atomic Energy Government of India (DAE), Department of Science and Technology, Government of India (DST), University Grants Commission, Government of India (UGC) and Council of Scientific and Industrial Research (CSIR), India;
National Research and Innovation Agency - BRIN, Indonesia;
Istituto Nazionale di Fisica Nucleare (INFN), Italy;
Japanese Ministry of Education, Culture, Sports, Science and Technology (MEXT) and Japan Society for the Promotion of Science (JSPS) KAKENHI, Japan;
Consejo Nacional de Ciencia (CONACYT) y Tecnolog\'{i}a, through Fondo de Cooperaci\'{o}n Internacional en Ciencia y Tecnolog\'{i}a (FONCICYT) and Direcci\'{o}n General de Asuntos del Personal Academico (DGAPA), Mexico;
Nederlandse Organisatie voor Wetenschappelijk Onderzoek (NWO), Netherlands;
The Research Council of Norway, Norway;
Pontificia Universidad Cat\'{o}lica del Per\'{u}, Peru;
Ministry of Science and Higher Education, National Science Centre and WUT ID-UB, Poland;
Korea Institute of Science and Technology Information and National Research Foundation of Korea (NRF), Republic of Korea;
Ministry of Education and Scientific Research, Institute of Atomic Physics, Ministry of Research and Innovation and Institute of Atomic Physics and Universitatea Nationala de Stiinta si Tehnologie Politehnica Bucuresti, Romania;
Ministerstvo skolstva, vyskumu, vyvoja a mladeze SR, Slovakia;
National Research Foundation of South Africa, South Africa;
Swedish Research Council (VR) and Knut \& Alice Wallenberg Foundation (KAW), Sweden;
European Organization for Nuclear Research, Switzerland;
Suranaree University of Technology (SUT), National Science and Technology Development Agency (NSTDA) and National Science, Research and Innovation Fund (NSRF via PMU-B B05F650021), Thailand;
Turkish Energy, Nuclear and Mineral Research Agency (TENMAK), Turkey;
National Academy of  Sciences of Ukraine, Ukraine;
Science and Technology Facilities Council (STFC), United Kingdom;
National Science Foundation of the United States of America (NSF) and United States Department of Energy, Office of Nuclear Physics (DOE NP), United States of America.
In addition, individual groups or members have received support from:
Czech Science Foundation (grant no. 23-07499S), Czech Republic;
FORTE project, reg.\ no.\ CZ.02.01.01/00/22\_008/0004632, Czech Republic, co-funded by the European Union, Czech Republic;
European Research Council (grant no. 950692), European Union;
Deutsche Forschungs Gemeinschaft (DFG, German Research Foundation) ``Neutrinos and Dark Matter in Astro- and Particle Physics'' (grant no. SFB 1258), Germany;
CONVECS project, CUP C97H23001700002 FESR 2021-2027 program, Italy;
ICSC - National Research Center for High Performance Computing, Big Data and Quantum Computing and FAIR - Future Artificial Intelligence Research, funded by the NextGenerationEU program (Italy).

\end{acknowledgement}

\bibliographystyle{utphys}   
\bibliography{bibliography}

\newpage
\appendix

%
%

\section{The ALICE Collaboration}
\label{app:collab}
\begin{flushleft} 
\small

S.~Acharya\,\orcidlink{0000-0002-9213-5329}\,$^{\rm 50}$, 
A.~Agarwal$^{\rm 133}$, 
G.~Aglieri Rinella\,\orcidlink{0000-0002-9611-3696}\,$^{\rm 32}$, 
L.~Aglietta\,\orcidlink{0009-0003-0763-6802}\,$^{\rm 24}$, 
M.~Agnello\,\orcidlink{0000-0002-0760-5075}\,$^{\rm 29}$, 
N.~Agrawal\,\orcidlink{0000-0003-0348-9836}\,$^{\rm 25}$, 
Z.~Ahammed\,\orcidlink{0000-0001-5241-7412}\,$^{\rm 133}$, 
S.~Ahmad\,\orcidlink{0000-0003-0497-5705}\,$^{\rm 15}$, 
S.U.~Ahn\,\orcidlink{0000-0001-8847-489X}\,$^{\rm 71}$, 
I.~Ahuja\,\orcidlink{0000-0002-4417-1392}\,$^{\rm 36}$
% ...

\section*{Affiliation Notes}

$^{\rm I}$ Deceased\\
$^{\rm II}$ Also at: Max-Planck-Institut fur Physik, Munich, Germany\\
$^{\rm III}$ Also at: Czech Technical University in Prague, Prague, Czech Republic\\
$^{\rm IV}$ Also at: Italian National Agency for New Technologies, Energy and Sustainable Economic Development (ENEA), Bologna, Italy\\
$^{\rm V}$ Also at: Instituto de Fisica da Universidade de Sao Paulo\\
$^{\rm VI}$ Also at: Dipartimento DET del Politecnico di Torino, Turin, Italy\\
$^{\rm VII}$ Also at: University College of Dublin, Dublin, Ireland\\
$^{\rm VIII}$ Also at: Department of Applied Physics, Aligarh Muslim University, Aligarh, India\\
$^{\rm IX}$ Also at: Institute of Theoretical Physics, University of Wroclaw, Wroclaw, Poland\\
$^{\rm X}$ Also at: Facultad de Ciencias, Universidad Nacional Aut\'{o}noma de M\'{e}xico, Mexico City, Mexico\\
$^{\rm XI}$ Also at: Lovely Professional University, Punjab, India
\\

\section*{Collaboration Institutes}

$^{1}$ A.I. Alikhanyan National Science Laboratory (Yerevan Physics Institute) Foundation, Yerevan, Armenia\\
$^{2}$ AGH University of Krakow, Cracow, Poland\\
$^{3}$ Bogolyubov Institute for Theoretical Physics, National Academy of Sciences of Ukraine, Kyiv, Ukraine\\
$^{4}$ Bose Institute, Department of Physics  and Centre for Astroparticle Physics and Space Science (CAPSS), Kolkata, India\\
$^{5}$ California Polytechnic State University, San Luis Obispo, California, United States\\
$^{6}$ Central China Normal University, Wuhan, China\\
$^{7}$ Centro de Aplicaciones Tecnol\'{o}gicas y Desarrollo Nuclear (CEADEN), Havana, Cuba\\
$^{8}$ Centro de Investigaci\'{o}n y de Estudios Avanzados (CINVESTAV), Mexico City and M\'{e}rida, Mexico\\
$^{9}$ Chicago State University, Chicago, Illinois, United States\\
$^{10}$ China Nuclear Data Center, China Institute of Atomic Energy, Beijing, China\\
$^{11}$ China University of Geosciences, Wuhan, China\\
$^{12}$ Chungbuk National University, Cheongju, Republic of Korea\\
$^{13}$ Comenius University Bratislava, Faculty of Mathematics, Physics and Informatics, Bratislava, Slovak Republic\\
$^{14}$ Creighton University, Omaha, Nebraska, United States\\
$^{15}$ Department of Physics, Aligarh Muslim University, Aligarh, India\\
$^{16}$ Department of Physics, Pusan National University, Pusan, Republic of Korea\\
$^{17}$ Department of Physics, Sejong University, Seoul, Republic of Korea\\
$^{18}$ Department of Physics, University of California, Berkeley, California, United States\\
$^{19}$ Department of Physics, University of Oslo, Oslo, Norway\\
$^{20}$ Department of Physics and Technology, University of Bergen, Bergen, Norway\\
$^{21}$ Dipartimento di Fisica, Universit\`{a} di Pavia, Pavia, Italy\\
$^{22}$ Dipartimento di Fisica dell'Universit\`{a} and Sezione INFN, Cagliari, Italy\\
$^{23}$ Dipartimento di Fisica dell'Universit\`{a} and Sezione INFN, Trieste, Italy\\
$^{24}$ Dipartimento di Fisica dell'Universit\`{a} and Sezione INFN, Turin, Italy\\
$^{25}$ Dipartimento di Fisica e Astronomia dell'Universit\`{a} and Sezione INFN, Bologna, Italy\\
$^{26}$ Dipartimento di Fisica e Astronomia dell'Universit\`{a} and Sezione INFN, Catania, Italy\\
$^{27}$ Dipartimento di Fisica e Astronomia dell'Universit\`{a} and Sezione INFN, Padova, Italy\\
$^{28}$ Dipartimento di Fisica `E.R.~Caianiello' dell'Universit\`{a} and Gruppo Collegato INFN, Salerno, Italy\\
$^{29}$ Dipartimento DISAT del Politecnico and Sezione INFN, Turin, Italy\\
$^{30}$ Dipartimento di Scienze MIFT, Universit\`{a} di Messina, Messina, Italy\\
$^{31}$ Dipartimento Interateneo di Fisica `M.~Merlin' and Sezione INFN, Bari, Italy\\
$^{32}$ European Organization for Nuclear Research (CERN), Geneva, Switzerland\\
$^{33}$ Faculty of Electrical Engineering, Mechanical Engineering and Naval Architecture, University of Split, Split, Croatia\\
$^{34}$ Faculty of Nuclear Sciences and Physical Engineering, Czech Technical University in Prague, Prague, Czech Republic\\
$^{35}$ Faculty of Physics, Sofia University, Sofia, Bulgaria\\
$^{36}$ Faculty of Science, P.J.~\v{S}af\'{a}rik University, Ko\v{s}ice, Slovak Republic\\
$^{37}$ Faculty of Technology, Environmental and Social Sciences, Bergen, Norway\\
$^{38}$ Frankfurt Institute for Advanced Studies, Johann Wolfgang Goethe-Universit\"{a}t Frankfurt, Frankfurt, Germany\\
$^{39}$ Fudan University, Shanghai, China\\
$^{40}$ Gangneung-Wonju National University, Gangneung, Republic of Korea\\
$^{41}$ Gauhati University, Department of Physics, Guwahati, India\\
$^{42}$ Helmholtz-Institut f\"{u}r Strahlen- und Kernphysik, Rheinische Friedrich-Wilhelms-Universit\"{a}t Bonn, Bonn, Germany\\
$^{43}$ Helsinki Institute of Physics (HIP), Helsinki, Finland\\
$^{44}$ High Energy Physics Group,  Universidad Aut\'{o}noma de Puebla, Puebla, Mexico\\
$^{45}$ Horia Hulubei National Institute of Physics and Nuclear Engineering, Bucharest, Romania\\
$^{46}$ HUN-REN Wigner Research Centre for Physics, Budapest, Hungary\\
$^{47}$ Indian Institute of Technology Bombay (IIT), Mumbai, India\\
$^{48}$ Indian Institute of Technology Indore, Indore, India\\
$^{49}$ INFN, Laboratori Nazionali di Frascati, Frascati, Italy\\
$^{50}$ INFN, Sezione di Bari, Bari, Italy\\
$^{51}$ INFN, Sezione di Bologna, Bologna, Italy\\
$^{52}$ INFN, Sezione di Cagliari, Cagliari, Italy\\
$^{53}$ INFN, Sezione di Catania, Catania, Italy\\
$^{54}$ INFN, Sezione di Padova, Padova, Italy\\
$^{55}$ INFN, Sezione di Pavia, Pavia, Italy\\
$^{56}$ INFN, Sezione di Torino, Turin, Italy\\
$^{57}$ INFN, Sezione di Trieste, Trieste, Italy\\
$^{58}$ Inha University, Incheon, Republic of Korea\\
$^{59}$ Institute for Gravitational and Subatomic Physics (GRASP), Utrecht University/Nikhef, Utrecht, Netherlands\\
$^{60}$ Institute of Experimental Physics, Slovak Academy of Sciences, Ko\v{s}ice, Slovak Republic\\
$^{61}$ Institute of Physics, Homi Bhabha National Institute, Bhubaneswar, India\\
$^{62}$ Institute of Physics of the Czech Academy of Sciences, Prague, Czech Republic\\
$^{63}$ Institute of Space Science (ISS), Bucharest, Romania\\
$^{64}$ Institut f\"{u}r Kernphysik, Johann Wolfgang Goethe-Universit\"{a}t Frankfurt, Frankfurt, Germany\\
$^{65}$ Instituto de Ciencias Nucleares, Universidad Nacional Aut\'{o}noma de M\'{e}xico, Mexico City, Mexico\\
$^{66}$ Instituto de F\'{i}sica, Universidade Federal do Rio Grande do Sul (UFRGS), Porto Alegre, Brazil\\
$^{67}$ Instituto de F\'{\i}sica, Universidad Nacional Aut\'{o}noma de M\'{e}xico, Mexico City, Mexico\\
$^{68}$ iThemba LABS, National Research Foundation, Somerset West, South Africa\\
$^{69}$ Jeonbuk National University, Jeonju, Republic of Korea\\
$^{70}$ Johann-Wolfgang-Goethe Universit\"{a}t Frankfurt Institut f\"{u}r Informatik, Fachbereich Informatik und Mathematik, Frankfurt, Germany\\
$^{71}$ Korea Institute of Science and Technology Information, Daejeon, Republic of Korea\\
$^{72}$ Laboratoire de Physique Subatomique et de Cosmologie, Universit\'{e} Grenoble-Alpes, CNRS-IN2P3, Grenoble, France\\
$^{73}$ Lawrence Berkeley National Laboratory, Berkeley, California, United States\\
$^{74}$ Lund University Department of Physics, Division of Particle Physics, Lund, Sweden\\
$^{75}$ Marietta Blau Institute, Vienna, Austria\\
$^{76}$ Nagasaki Institute of Applied Science, Nagasaki, Japan\\
$^{77}$ Nara Women{'}s University (NWU), Nara, Japan\\
$^{78}$ National and Kapodistrian University of Athens, School of Science, Department of Physics , Athens, Greece\\
$^{79}$ National Centre for Nuclear Research, Warsaw, Poland\\
$^{80}$ National Institute of Science Education and Research, Homi Bhabha National Institute, Jatni, India\\
$^{81}$ National Nuclear Research Center, Baku, Azerbaijan\\
$^{82}$ National Research and Innovation Agency - BRIN, Jakarta, Indonesia\\
$^{83}$ Niels Bohr Institute, University of Copenhagen, Copenhagen, Denmark\\
$^{84}$ Nikhef, National institute for subatomic physics, Amsterdam, Netherlands\\
$^{85}$ Nuclear Physics Group, STFC Daresbury Laboratory, Daresbury, United Kingdom\\
$^{86}$ Nuclear Physics Institute of the Czech Academy of Sciences, Husinec-\v{R}e\v{z}, Czech Republic\\
$^{87}$ Oak Ridge National Laboratory, Oak Ridge, Tennessee, United States\\
$^{88}$ Ohio State University, Columbus, Ohio, United States\\
$^{89}$ Physics Department, Panjab University, Chandigarh, India\\
$^{90}$ Physics Department, University of Jammu, Jammu, India\\
$^{91}$ Physics Program and International Institute for Sustainability with Knotted Chiral Meta Matter (WPI-SKCM$^{2}$), Hiroshima University, Hiroshima, Japan\\
$^{92}$ Physikalisches Institut, Eberhard-Karls-Universit\"{a}t T\"{u}bingen, T\"{u}bingen, Germany\\
$^{93}$ Physikalisches Institut, Ruprecht-Karls-Universit\"{a}t Heidelberg, Heidelberg, Germany\\
$^{94}$ Physik Department, Technische Universit\"{a}t M\"{u}nchen, Munich, Germany\\
$^{95}$ Politecnico di Bari and Sezione INFN, Bari, Italy\\
$^{96}$ Research Division and ExtreMe Matter Institute EMMI, GSI Helmholtzzentrum f\"ur Schwerionenforschung GmbH, Darmstadt, Germany\\
$^{97}$ Saga University, Saga, Japan\\
$^{98}$ Saha Institute of Nuclear Physics, Homi Bhabha National Institute, Kolkata, India\\
$^{99}$ School of Physics and Astronomy, University of Birmingham, Birmingham, United Kingdom\\
$^{100}$ Secci\'{o}n F\'{\i}sica, Departamento de Ciencias, Pontificia Universidad Cat\'{o}lica del Per\'{u}, Lima, Peru\\
$^{101}$ SUBATECH, IMT Atlantique, Nantes Universit\'{e}, CNRS-IN2P3, Nantes, France\\
$^{102}$ Sungkyunkwan University, Suwon City, Republic of Korea\\
$^{103}$ Suranaree University of Technology, Nakhon Ratchasima, Thailand\\
$^{104}$ Technical University of Ko\v{s}ice, Ko\v{s}ice, Slovak Republic\\
$^{105}$ The Henryk Niewodniczanski Institute of Nuclear Physics, Polish Academy of Sciences, Cracow, Poland\\
$^{106}$ The University of Texas at Austin, Austin, Texas, United States\\
$^{107}$ Universidad Aut\'{o}noma de Sinaloa, Culiac\'{a}n, Mexico\\
$^{108}$ Universidade de S\~{a}o Paulo (USP), S\~{a}o Paulo, Brazil\\
$^{109}$ Universidade Estadual de Campinas (UNICAMP), Campinas, Brazil\\
$^{110}$ Universidade Federal do ABC, Santo Andre, Brazil\\
$^{111}$ Universitatea Nationala de Stiinta si Tehnologie Politehnica Bucuresti, Bucharest, Romania\\
$^{112}$ University of Derby, Derby, United Kingdom\\
$^{113}$ University of Houston, Houston, Texas, United States\\
$^{114}$ University of Jyv\"{a}skyl\"{a}, Jyv\"{a}skyl\"{a}, Finland\\
$^{115}$ University of Kansas, Lawrence, Kansas, United States\\
$^{116}$ University of Liverpool, Liverpool, United Kingdom\\
$^{117}$ University of Science and Technology of China, Hefei, China\\
$^{118}$ University of South-Eastern Norway, Kongsberg, Norway\\
$^{119}$ University of Tennessee, Knoxville, Tennessee, United States\\
$^{120}$ University of the Witwatersrand, Johannesburg, South Africa\\
$^{121}$ University of Tokyo, Tokyo, Japan\\
$^{122}$ University of Tsukuba, Tsukuba, Japan\\
$^{123}$ University of Zagreb Faculty of Science, Department of Physics, Zagreb, Croatia\\
$^{124}$ Universit\"{a}t M\"{u}nster, Institut f\"{u}r Kernphysik, M\"{u}nster, Germany\\
$^{125}$ Universit\'{e} Clermont Auvergne, CNRS/IN2P3, LPC, Clermont-Ferrand, France\\
$^{126}$ Universit\'{e} de Lyon, CNRS/IN2P3, Institut de Physique des 2 Infinis de Lyon, Lyon, France\\
$^{127}$ Universit\'{e} de Strasbourg, CNRS, IPHC UMR 7178, F-67000 Strasbourg, France, Strasbourg, France\\
$^{128}$ Universit\'{e} Paris-Saclay, Centre d'Etudes de Saclay (CEA), IRFU, D\'{e}partment de Physique Nucl\'{e}aire (DPhN), Saclay, France\\
$^{129}$ Universit\'{e}  Paris-Saclay, CNRS/IN2P3, IJCLab, Orsay, France\\
$^{130}$ Universit\`{a} degli Studi di Foggia, Foggia, Italy\\
$^{131}$ Universit\`{a} del Piemonte Orientale, Vercelli, Italy\\
$^{132}$ Universit\`{a} di Brescia, Brescia, Italy\\
$^{133}$ Variable Energy Cyclotron Centre, Homi Bhabha National Institute, Kolkata, India\\
$^{134}$ Warsaw University of Technology, Warsaw, Poland\\
$^{135}$ Wayne State University, Detroit, Michigan, United States\\
$^{136}$ Yale University, New Haven, Connecticut, United States\\
$^{137}$ Yildiz Technical University, Istanbul, Turkey\\
$^{138}$ Yonsei University, Seoul, Republic of Korea\\
$^{139}$ Affiliated with an institute formerly covered by a cooperation agreement with CERN\\
$^{140}$ Affiliated with an international laboratory covered by a cooperation agreement with CERN.\\

\end{flushleft} 

\end{document}